\documentclass[12pt,preprint]{aastex}

\shorttitle{The Nature of Ultra Luminous X-ray Sources}
\begin{document}
\title{The Nature of Ultra Luminous X-ray Sources}
\author{C. M. Guti\'errez$^1$, and M. L\'opez-Corredoira$^{1}$}
\affil{\small$^1$Instituto de Astrof\'{\i}sica de Canarias, E-38205, La Laguna, 
Tenerife, Spain (cgc@ll.iac.es, martinlc@iac.es)}

\begin{abstract}

We present spectroscopic observations of six
optical counterparts of intermediate luminosity
X-ray  sources (ULXs) around nearby galaxies. 
The spectra of the six objects show  the presence
of  broad emission features. The identification
of these allow us to classify all of the objects 
as quasars at higher redshift than their assigned
parent galaxy. This is one of the first and
largest identifications of such objects using
unambigous optical spectral features. These
results, in conjuction with previous  similar
identifications of other sources,  indicate that
high redshift quasars represent an important
fraction of  catalogued ULX sources. We estimate
the density of such sources and compare this with
expectations for a population of randomly
distributed background quasars. 

\end{abstract}

\keywords{galaxies: active - quasars: general - X-rays: galaxies}






\section{Introduction}

One of the most intriguing astrophysical objects are the ultraluminous X-ray sources
(ULXs) discovered around nearby galaxies
by the X-ray satellites {\itshape Einstein}, {\itshape  ROSAT}, {\itshape Chandra} and {\itshape XMM--Newton}.
Assuming that they are at the same distance as the 
galaxies that are apparently close to them on the sky, their luminosities in the 0.1--2.4 keV X-ray band are in the range
$10^{39}$--$10^{40}$ erg s$^{-1}$, so that they are too luminous to be X-ray
cataclysmic binary stars (Pakull \& Mirioni 2002) and too faint  to be compared with the
typical luminosities of  active galactic nuclei.  
 Although much remains to be clarified, several
explanations concerning the nature of these objects in terms of globular clusters, HII
regions, etc.\ (Pakull \& Mirioni 2002; Angelini, Loewenstein \& Mushotzky 2001; Gao
et al.\ 2003; Roberts, Goad, \& Warwick 2003; Wang 2002), or  of hypothetical
supermassive stars or beamed emission (King et al.\ 2001; K\"ording, Falcke, \&
Markoff 2002) have been proposed. Studies with {\itshape XMM--Newton}
(Jenkins et al.\ 2004)  point to a  heterogeneus class of objects whose spectral
properties are similar to those of objects with lower X-ray luminosities.

A catalogue of ULXs around bright nearby galaxies has been compiled (Colbert \& Ptak
2002, CP02 hereafter) by cross-correlating public archive {\itshape ROSAT} HRI images with a selection of
nearby ($cz\le 5000$ km s$^{-1}$) galaxies taken from the RC3 catalogue (de
Vaucouleurs et al.\ 1991). The catalogue of ULXs  (or IXOs, Intermediate X-ray
Objects, in the notation of CP02) comprises 87 objects  around 54 bright
nearby  galaxies.  According to the analysis of the CP02 catalogue made
by Irwin, Bregman \& Athey (2004) ULXs with luminosities below  $\sim 2\times
10^{39}$ erg s$^{-1}$ can be explained naturally by accretion on to black holes with
masses $\sim 10$--20$ M_o$. The analysis of objects  with higher X-ray luminosities has
revealed that only for parent galaxies of late types is the distribution of objects 
concentrated towards the galactic centre. Also, according to the analysis of these authors,
the number of ULXs is smaller in
early type parent galaxies and compatible with the expected number of
foreground/background sources, while  the
number of ULXs around late-type galaxies is larger  and  requires the existence of a
new class of objects within the parent galaxy. The contamination of background
objects obviously increases with distance to the parent galaxy; in fact Ptak \&
Colbert (2004) have claimed that most of the ULXs listed in the CP02 catalogue  at distances
from the assumed parent galaxy $\ge 0.5D_{25}$ are probably background objects. 

These considerations on the nature of ULXs are based on indirect statistical
evidence. The confirmation of this evidence with the identification of counterparts in other
bands is essential. The number of such optical 
identifications is still small (e.g.\ Roberts et al.\ 2001; Wu et al.\ 2002; Maseti et al.\ 2003; Liu,
Bregman \& Seitzer 2004) and  is usually restricted to single galaxies, making
statistical studies very difficult. The study presented here includes sources in
several galaxies and allows one of the first  identifications based
on unambiguous spectral features in the optical. We have looked for these
counterparts of the ULX sources listed in the CP02 catalogue  in the Digital Sky
Survey (DSS) images and in the digitalized USNO catalogues. The typical magnitudes of such objects are
18--20 mag in the $b$ band and are therefore bright enough targets  for spectroscopic
observations with 2 to 4 m telescopes. Previously we have presented (Arp,
Guti\'errez \& L\'opez-Corredoira 2004) an analysis of the first three objects
observed; two of them (IXO~1 and 2), close to the galaxy NGC~720, resulted in quasars
at much higher redshift  (0.959 and 2.216 respectively) than the parent galaxy.
Given the paucity of quasars in the sky, a misidentification is unlikely. The third
one (IXO~5) was associated with an HII region at the redshift of the parent galaxy
(NGC~1073).  This last object was at the edge of a spiral arm of the parent galaxy,
where there are many HII regions. If the identification is correct, these would confirm
previous findings (Gao et al.\ 2003) claiming that many of the X-ray sources around 
the Cartwheel galaxy are associated with HII regions. In this paper, we present the
results of new observations of six other cases, all of which also turn out to be quasars at
higher redshift than their parent galaxies. 

\section{Observations and data analysis}

We took spectroscopy in January 2004 for the optical counterparts of sources (following the notation of the
CP02 catalogue)
IXO~33, 45, 58, 69, 71 and 84. The observations were conducted at the William Herschel Telescope 
(WHT)\footnote{The William Herschel Telescope  is operated by the Isaac Newton
Group and the IAC in Spain's Roque de los Muchachos Observatory. The  observations
were done in service time.}. We used the red arm of the ISIS spectrograph with the
R158R grism. The slit width was between 1.20 and 1.45  arcsec. We  used Cu--Ar and
Cu--Ne lamps for wavelength calibration. This  provides a sampling of 1.62 \AA\,
pixel$^{-1}$ and an effective resolution of 8--10 \AA~(depending on the slit width
used and  atmospheric seeing).  A single image with an exposure
time of 1800 s was taken for each object.  The spectra were analysed following a standard
procedure using IRAF\footnote{IRAF is the Image Reduction and Analysis Facility,
written and supported by the IRAF programming group at the national Optical
Astronomy Observatories (NOAO) in Tucson, Arizona.} that comprises bias
subtraction, extraction of the spectra and wavelength calibration. We used the
standard spectroscopic stars  Feige~67 (Oke 1990), and BD+26 2606 (Oke
\& Gunn 1983) to  correct approximately for the response of the configuration to different
wavelengths. Given the prohibitive time needed to obtain flat-field images (especially
in the blue part of the spectrum), we did not correct for that effect. However, we
have checked that this correction would have been very small ($\leq 1$ \%) and would not  have
affected any of the  identifications of the main spectral features reported here.

Figure~1 shows the DSS images with the identification of the optical
counterpart. Figure~2 presents the spectra taken for these objects. Spectra of
four of the objects (IXO~45, 58, 69 and 71) are rather flat while the other two
(IXO~33 and 84) rise slightly towards red wavelengths. The main lines in the
spectra of objects  IXO~33 and 71 are wide Balmer lines (H$_\alpha$ and
H$_\beta$) and  narrow forbidden OII ($\lambda \lambda 3727$ \AA) and OIII
($\lambda \lambda 4959, \, 5007$ \AA) lines, while the spectra of the other
four (objects IXO~45, 58, 69 and 84) are dominated by  the wide emission line
of Mg II (2798 \AA).  All these features are characteristic of quasar spectra.
The position of the main emission lines allows  unambigous  determination of
the redshifts of the objects. The uncertainties in the determination of
redshift are $\sim$0.001 for objects in which we detect the MgII lines and
$\sim$0.0002 for the rest. Table~1 presents a summary of the main properties of
the objects and the estimated redshifts. 

{\bf Other objects in the slit}

In three of the cases (IXO~33, 58 and 84) there is more than one
object in the DSS very close to the main object and whose position
is  marginally compatible with that of the X-ray source. In all these cases we
positioned the slit crossing  these objects also.  These are:

\begin{itemize}

\item Second object near IXO~33: This object, not listed in the USNO catalogue,
is nevertheless visible in the DSS plates. It is at a distance $\sim18$ arcsec
from the main object. The spectra has  absorption features typical of a local
cold star. 

\item Second object near IXO~58: According to the USNO catalogue, this object
has 18.8 and 18.5 mag in $b$ and $r$ bands respectively and  is at a distance of $\sim$14  arcsec 
from the main object. We have identified weak Balmer H$_\delta$,
H$_\gamma$,  H$_\beta$ and H$_\alpha$ lines at z=0.317 in its continuum, which has no strong features. We tentatively
identify the spectra as a BL Lac type object.

\item  Objects near IXO~84: An object at $\sim$18 arcsec from the main one  
has a very weak continuum in which we detect a narrow emission line at 6843
\AA, which we were unable to identify. By chance, the slit also crosses
another  object at a distance of $\sim$64  arcsec from the main object. It
has a spectum typical of an HII region resulting a redshift of $z=0.0525$. The
object is extended and, given its redshift and its location near an arm of the
galaxy  NGC~5774, it seems to be an HII region of this galaxy.

\end{itemize}

\section{Discussion and conclusions}

In the following analysis we also include  the three objects IXO~1, 2 and 5 that were observed with the
same set-up and presented in Arp et al.\ (2004).  
The ULXs analysed here have X-ray luminosities in the range  
$\sim$1--6$\times 10^{39}$  erg s$^{-1}$ (assuming 
that the X-ray sources are at the distance of the putative parent galaxy) and  are at projected  distances
in  the range $\sim$0.7--1.9$\times D_{25}$ from the 
centre of such galaxiex, six of them  in elliptical galaxies and three in spirals. All except for one (IXO~5)  
turn out to be quasars at higher redshift than the assumed parent galaxy.
These results extend the findings of Foschini et al.\ (2002) and  Maseti
et al.\ (2003), who also found  optical counterparts of ULXs at a higher
redshift than that of the assigned host galaxy, and  that other objects  obeying the 
observational criteria used to define ULXs were not included in the CP02
compilation because they were previously known  high redshift quasars
(see the discussion on this in Arp et al.\ 2004). Because our targets were selected
among those visible in the DSS, it should be reasonable to 
deduce that an important fraction (the majority in the case of early-type parent galaxies) of
the ULX sources with optical counterparts in the range $\sim$18--20 mag must
also be quasars. The only object that has a redshift similar to that of the
assigned parent galaxy is IXO~5, which is in a star forming region of 
the spiral  galaxy NGC~1073. 

It is interesting to compare these results with the statistics of quasars in the
field. This analysis is difficult because of the characteristics in terms of
completeness, coverage, etc., of the CP02 catalogue and the small size of our
sample.  A simplistic  and rough calculation can be made by considering the optical
magnitudes of the quasars detected, and the area surveyed by CP02. The total area
enclosed in a circle centred on the parent galaxy and enclosing the ULXs found in
our sample is $\sim$500 square arcmin. According to the study by Ptak \& Colbert
(2004) only $\sim$10 \% of the RC3 galaxies observed with HRI have at least one ULX,
so that the total area surveyed in which the quasars have been found is $\sim$1.5
square degrees. From the 2dF Quasar Survey (Croom et al.\ 2004) and the fit presented
by L\'opez-Corredoira \& Guti\'errez (2004) to the Boyle et al.\ (2000) data, there
are about 0.2, 3, and 17  quasars per square degree brighter than 18, 19 and 20 mag,
respectively, in the $b$ band. So, in principle, the inferred distribution of quasars
from our measurements is  roughly in  agreement with  expectations for a purely
random distribution of background quasars. This would confirm the statistical analysis by Irwin et al.\
(2004) and Ptak \& Colbert (2004), who claimed that the distribution  of ULXs around
early-type galaxies is compatible with  expectations from the  background
population of objects. However, an
accurate assesment of this would require a direct
comparison with the density of X-ray selected
background quasars. In any case,   searching for optical  counterparts of ULXs and taking their
spectra seems to be one of the most promising and  unambiguous methods of unravelling the
true nature of such objects.

\section*{Acknowledgements}

We thank H. Arp for fruitful discussions. We also thank the anonymous referees for their comments, 
and
important issues regarding the statistical analysis.

\newpage

\begin{figure}
\includegraphics[angle=0,scale=.38]{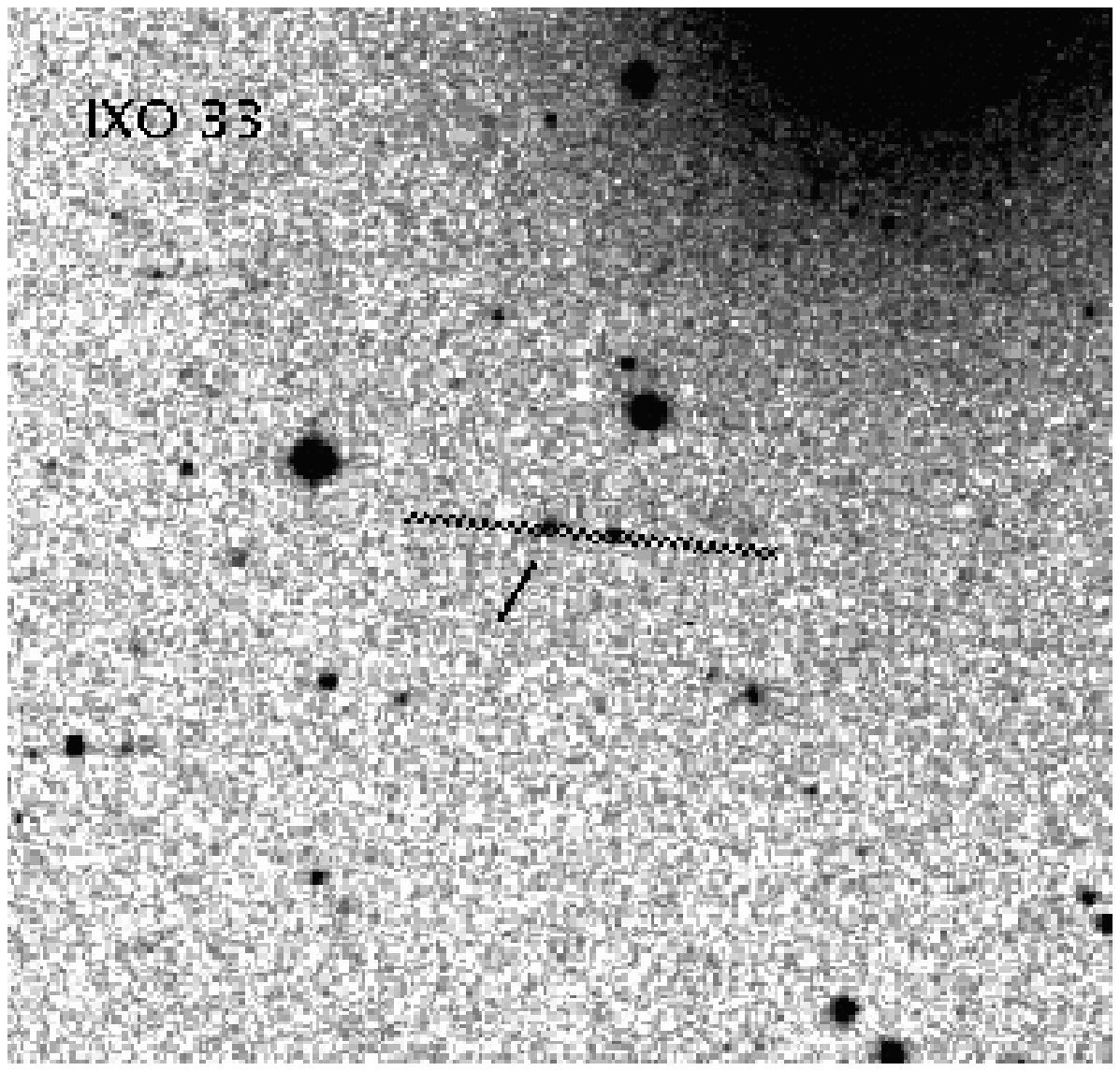}%
\includegraphics[angle=0,scale=.38]{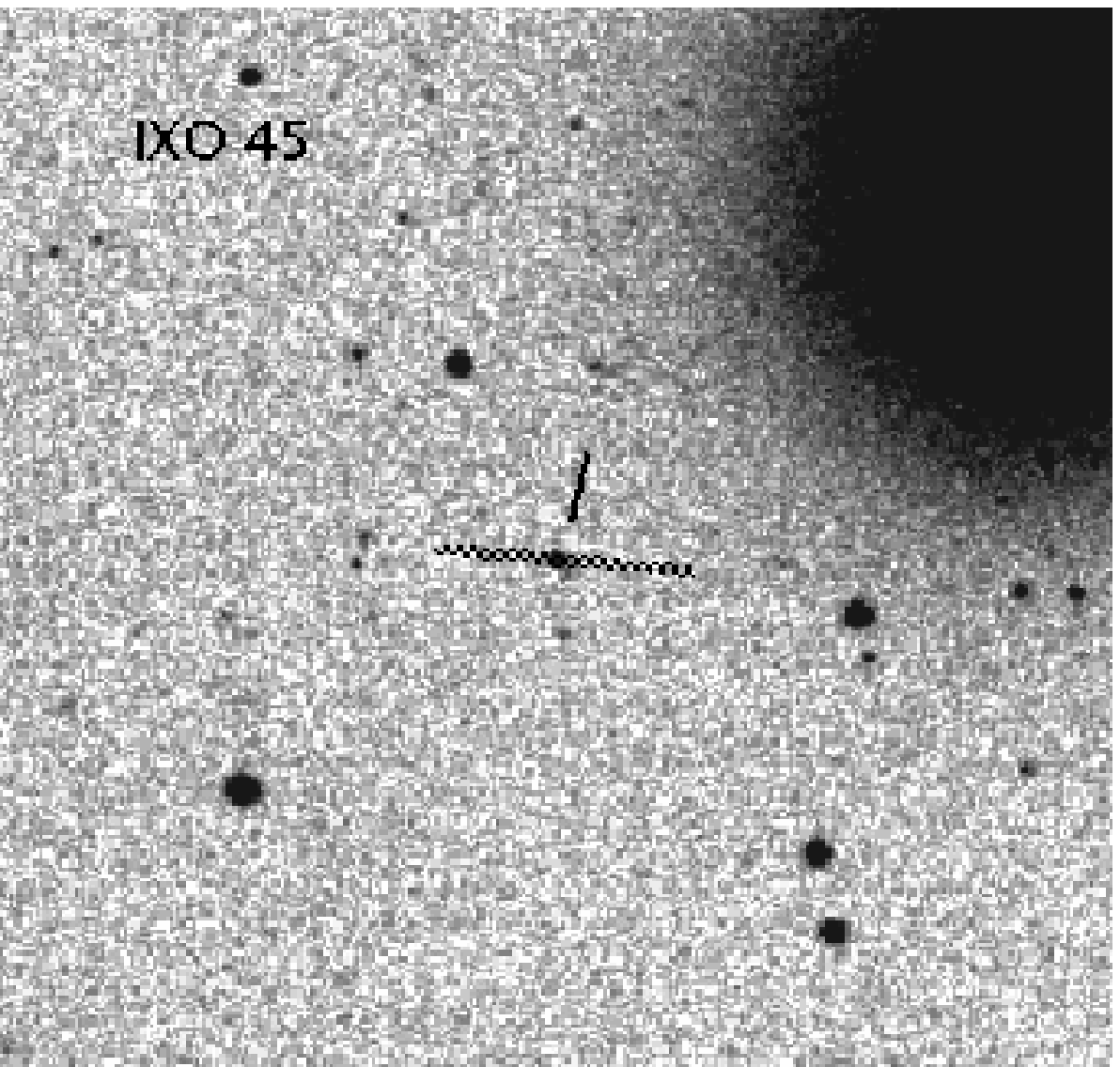}%
\includegraphics[angle=0,scale=.38]{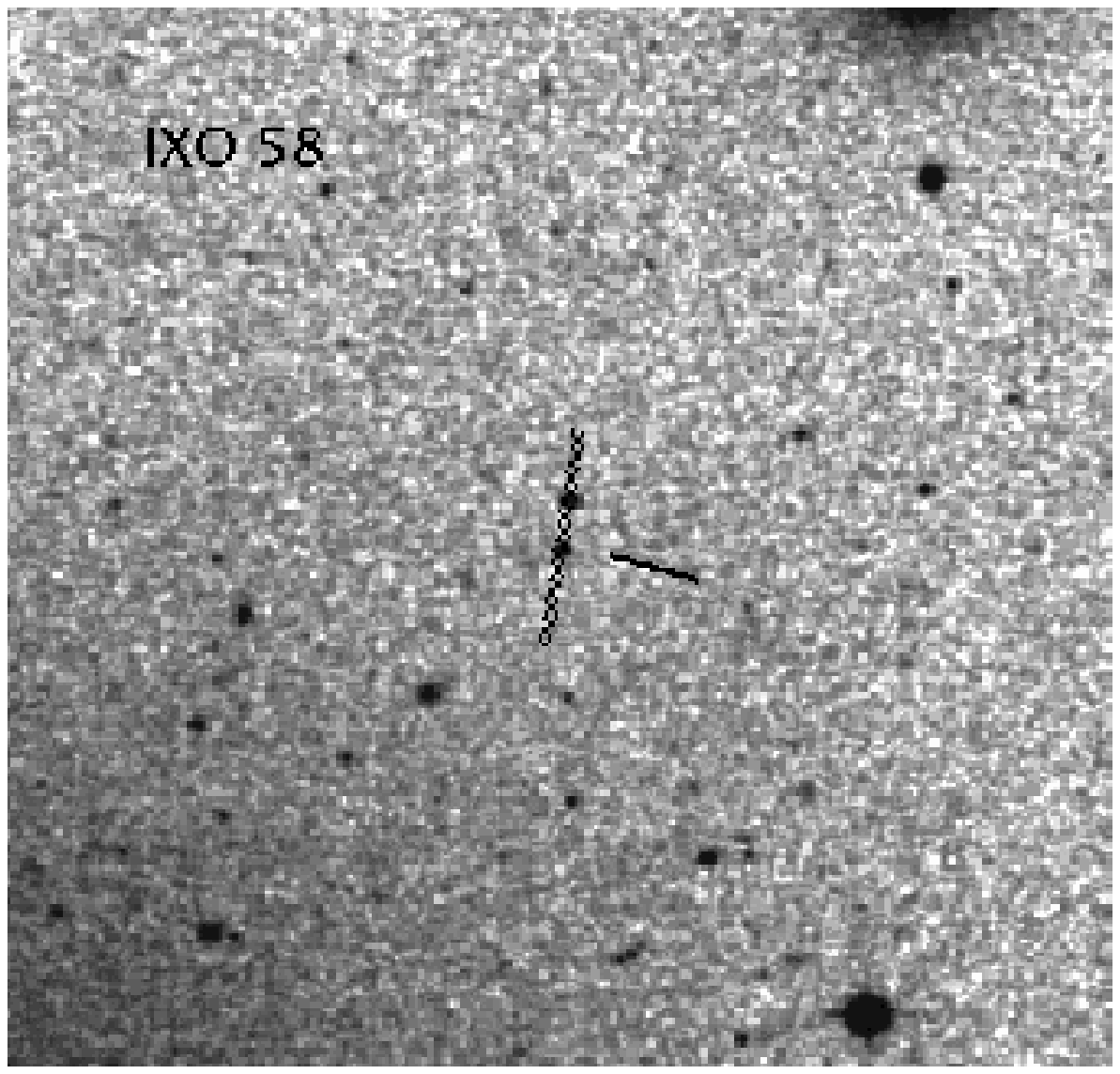}
\includegraphics[angle=0,scale=.38]{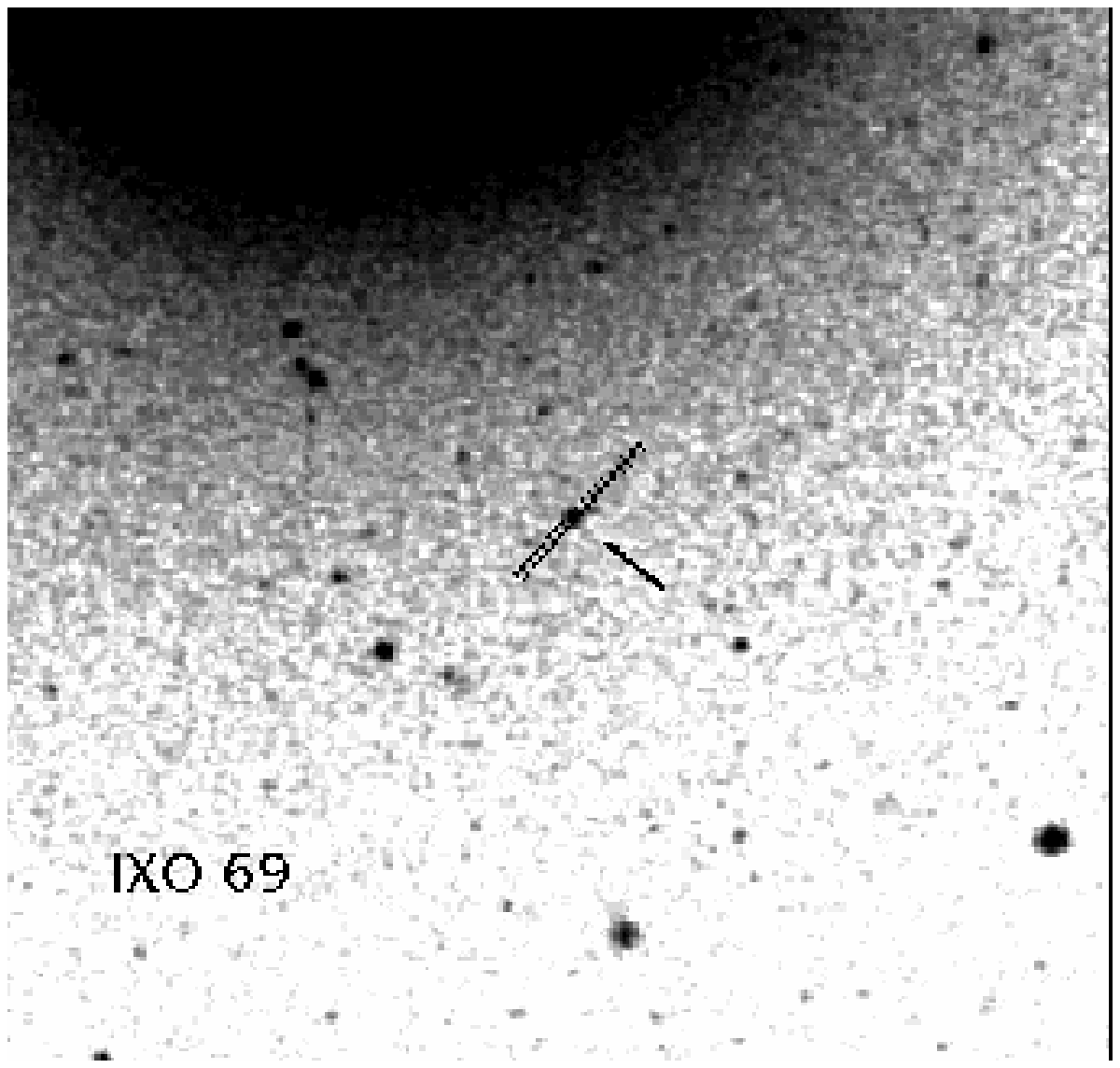}%
\includegraphics[angle=0,scale=.38]{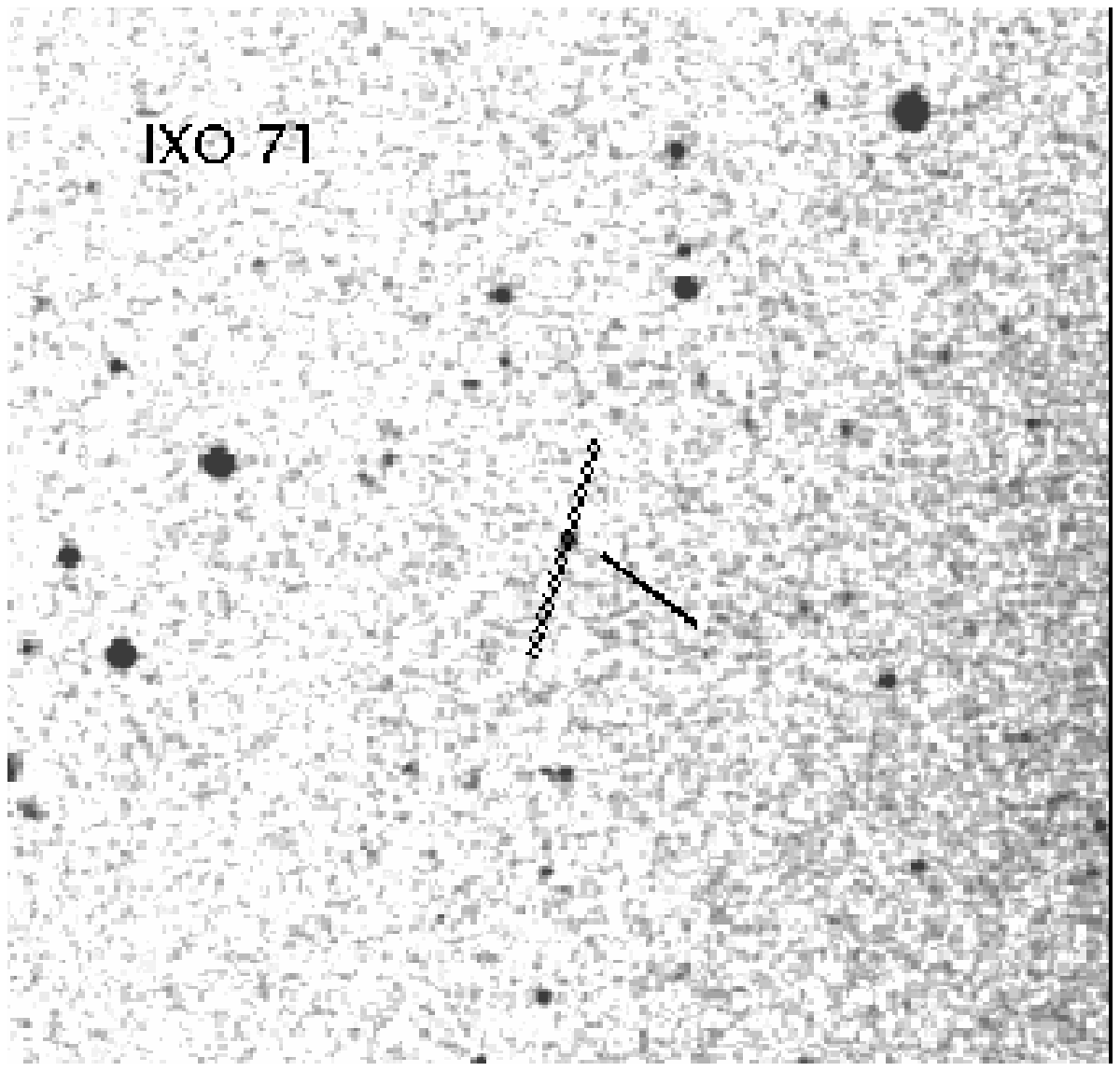}%
\includegraphics[angle=0,scale=.38]{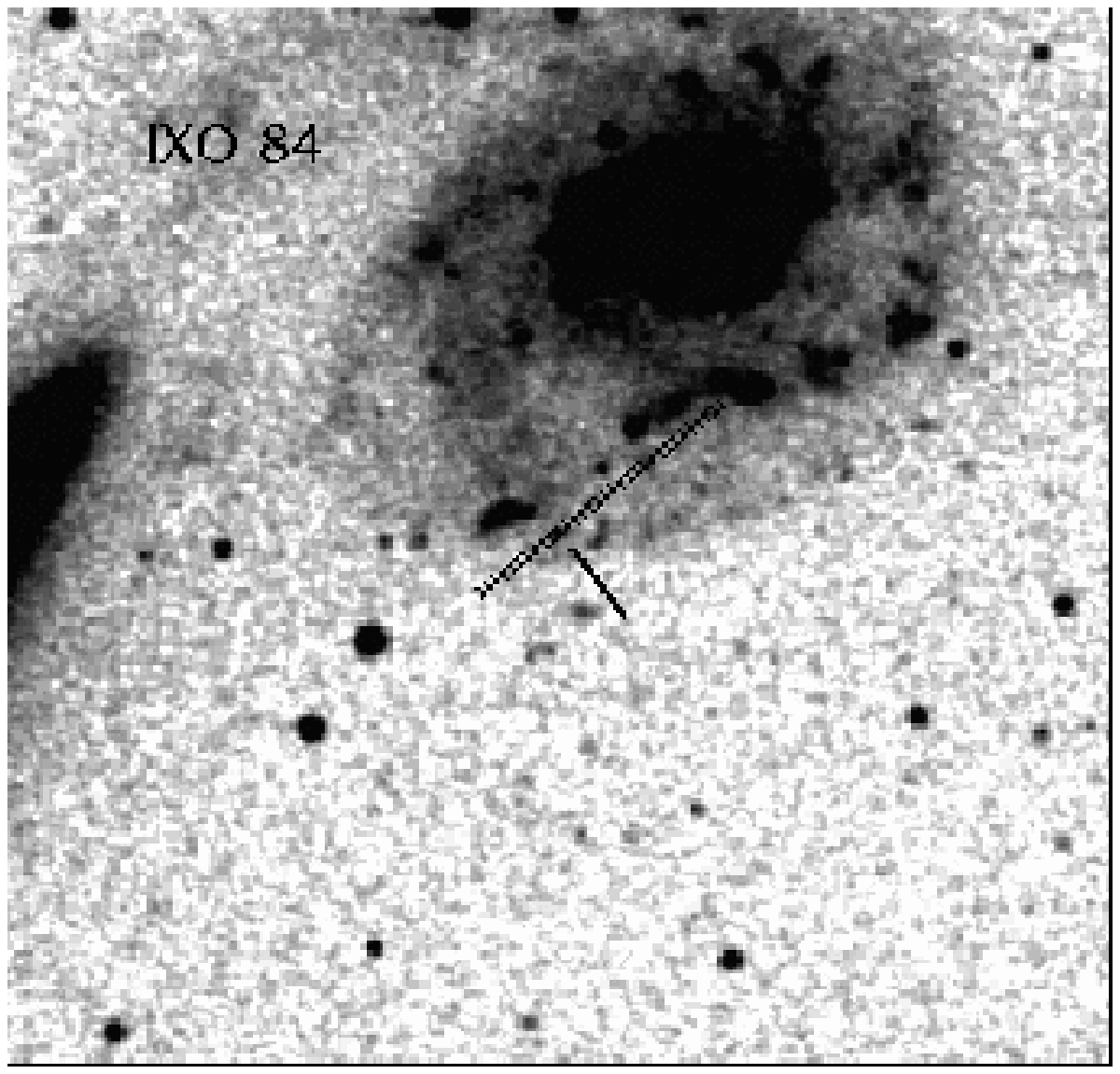}%
\vspace{7mm}
\caption{Images of the Digital
Sky Survey in the blue filter of 1 square arcminute centred on optical
counterparts of the ultraluminous X-ray sources analized in this paper. Names
according to the notation by Colbert \& Ptak (2002) are indicated. North is up
and East to the left. The small dark line indicate the position of such
objects. The grey wide lines indicate the orientation of the slit.}
\end{figure}

\newpage

\begin{figure} \caption{Optical spectra (for clarity, corrected by redshift) of the
sources marked on Figure~1. The y-axis is the flux in arbitrary units. The 
 main spectral features and the redshifts estimated for each object are indicated  The
gaps in the spectra correspond to the position of the telluric A band.}  
\includegraphics[angle=0,scale=.60]{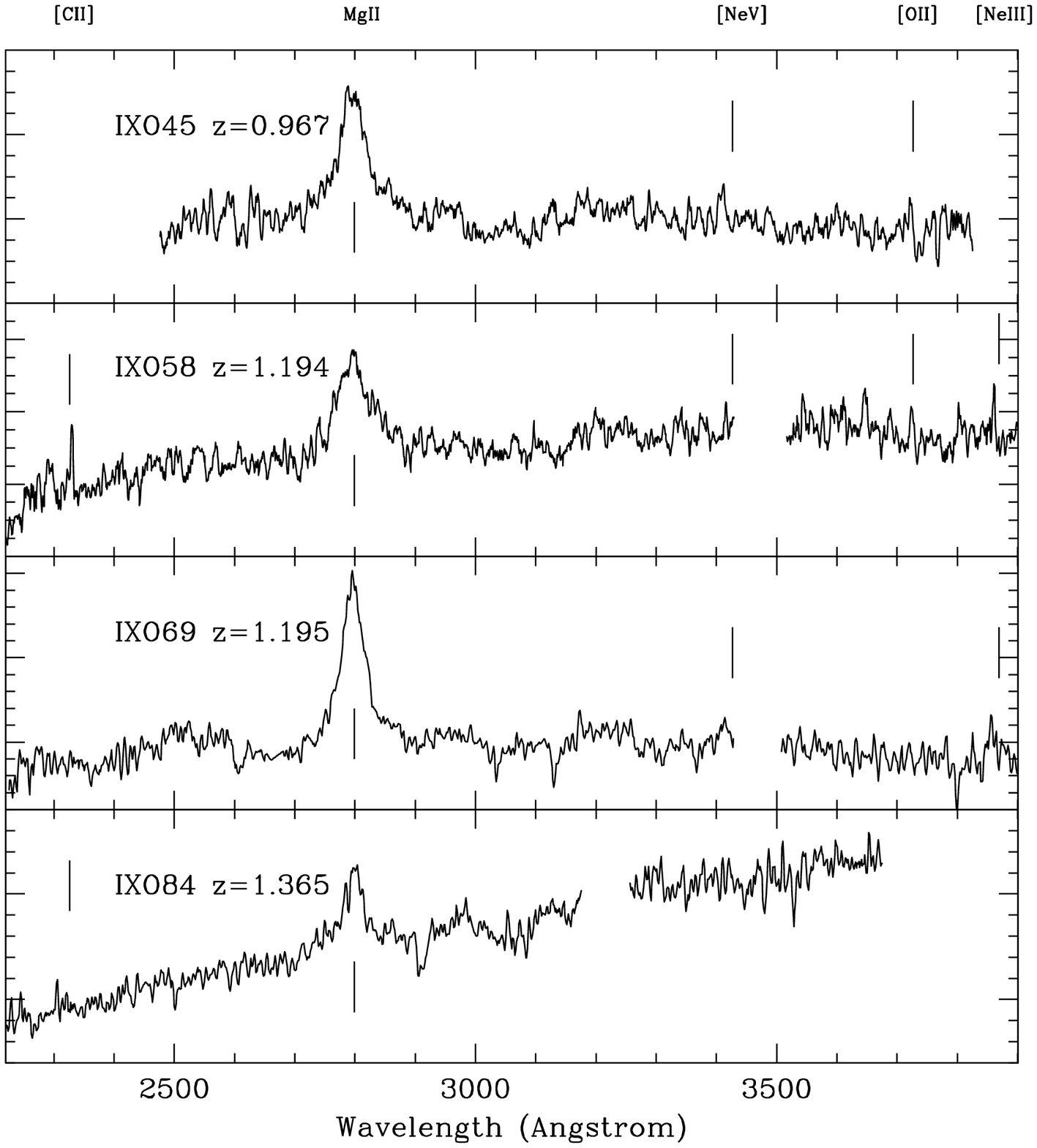}
\includegraphics[angle=0,scale=.60]{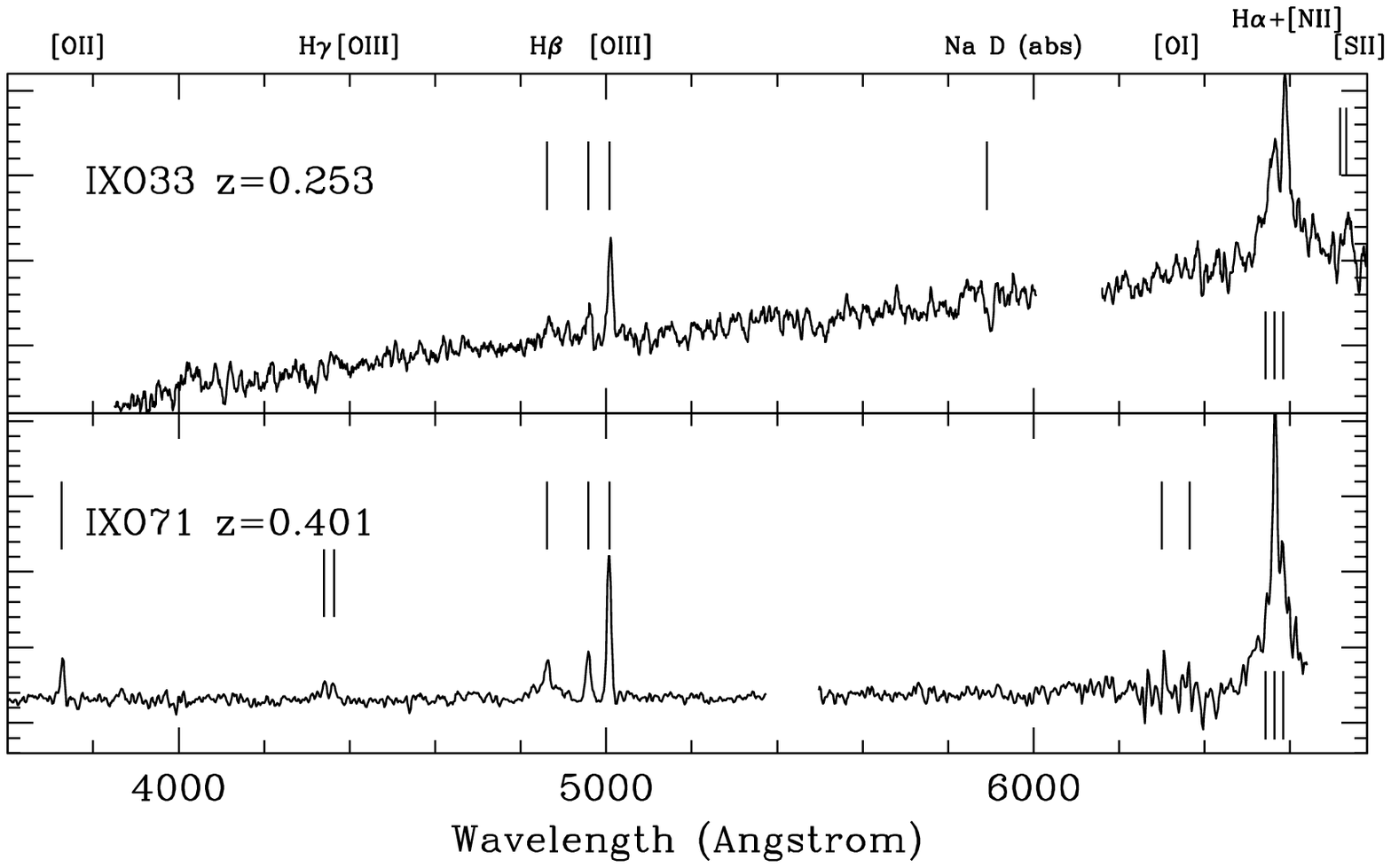}
\end{figure}

\newpage
\clearpage

\begin{deluxetable}{lccclccccc}
\tablewidth{0pt}
\tabletypesize{\scriptsize}
\tablecaption{Optical counterpart of ULX sources}
\tablehead{
\colhead{ID} & \colhead{RA (J2000)}  & \colhead{Dec (J2000)}  & \colhead{$\log \,(L_X)$}     
& \colhead{ID Gal.}   &
\colhead{$d$} & \colhead{$d/R_{25}$} & \colhead{$z_g$} & \colhead{$z$} & Object\\
 & \colhead{(hh:mm:ss.s)} 
 & \colhead{($^\circ$:$^\prime$:$^{\prime \prime}$)} & \colhead{(erg
 s$^{-1}$)} & 
 &  \colhead{(arcmin)} 
}
\startdata
IXO~1  & 01:52:49.7 & $-$13:42:11  & 39.1 & NGC~720  & 3.0 & 1.3 & 0.0058 & 2.216 & QSO\\
IXO~2  & 01:53:02.9 & $-$13:46:53  & 39.7 & NGC~720  & 2.6 & 1.1 & 0.0058 & 0.959 & QSO\\
IXO~5  & 02:43:38.3 &   +01:24:13  & 39.1 & NGC~1073 & 1.8 & 0.7 & 0.0040 & 0.0037& HII\\
IXO~33 & 09:10:27.0 &   +06:59:10  & 39.5 & NGC~2775 & 3.6 & 1.7 & 0.0045 & 0.253& QSO\\
IXO~45 & 12:15:15.7 &   +33:10:20  & 39.0 & NGC~4203 & 2.7 & 1.6 & 0.0036 & 0.967& QSO\\
IXO~58 & 12:29:27.9 &   +08:06:34  & 39.4 & NGC~4472 & 8.1 & 1.6 & 0.0033 & 1.194& QSO\\
IXO~69 & 12:43:36.8 &   +11:30:06  & 39.4 & NGC~4649 & 3.2 & 0.8 & 0.0037 & 1.195& QSO\\
IXO~71 & 12:44:09.2 &   +11:33:36  & 39.7 & NGC~4649 & 7.1 & 1.9 & 0.0037 & 0.401& QSO\\
IXO~84 & 14:53:44.7 &   +03:33:30  & 39.4 & NGC~5775 & 3.3 & 1.5 & 0.0056 & 1.365& QSO\\
\enddata
\tablenotetext{*}{
1. Identification of the ULXs following the Colbert \& Ptak (2002)
notation;
2-3. RA (J2000) and Dec (2000) position of the optical counterpart
according to the DSS plates and USNO catalogues; 
4. Luminosity in the band 0.1-2.4 KeV (in units of erg s$^{-1}$) assuming 
that the X-ray sources are at the distance of the parent galaxy; 
5. Identification of the parent galaxy; 
6-7. Angular distance (in arcmin and in units of $R_{25}$ of the central galaxy) between 
these galaxies and the X-ray source;
8 Redshift of the parent galaxy (taken from the literature); 
9. Redshift of the X-ray source as computed in this work. 10. Type of the optical counterpart.
The analysys of objects IXO~1, IXO~2 and IXO~5 was presented in Arp et al. (2004).
}
\end{deluxetable}

\end{document}